# Investigating introductory and advanced students' difficulties with entropy and the second law of thermodynamics using a validated instrument

Mary Jane Brundage[1], David E. Meltzer [2], and Chandralekha Singh[1]

[1] *Department of Physics and Astronomy, University of Pittsburgh, Pittsburgh, PA, 15260*
[2] *College of Integrative Sciences and Arts, Arizona State University, Mesa, AZ 85212*

We use the Survey of Thermodynamic Processes and First and Second Laws-Long (STPFaSL-Long), a research-based survey instrument with 78 items at the level of introductory physics, to investigate introductory and advanced students' difficulties with entropy and the second law of thermodynamics. We present analysis of data from 12 different introductory and advanced physics classes at four different higher education public institutions in the US in which the survey was administered in-person to more than 1000 students. We find that a widespread unproductive tendency for introductory students to associate the properties of entropy with those of energy leads to many errors based on an idea of "conservation of entropy," in which entropy increases are *always* balanced by equal entropy decreases. For many of the more advanced students (calculus-based and upper level), we detect a tendency to expect entropy increases even in processes in which the entropy does *not* change. We observed a widespread failure to correctly apply the relationship $\Delta S = \delta Q_{reversible}/T$, either by using it for *irreversible* processes to which it does not apply, or by applying it incorrectly or completely neglecting it in reversible processes to which it does apply. We also noted that many introductory students are simply not aware that total entropy must increase in any "spontaneous" heat transfer process. Students at all levels were very frequently found to be confused that while net entropy (system + reservoir) in reversible isothermal processes does not change, the entropy of the working substance itself does indeed increase or decrease depending on whether the process is an expansion or compression. Our findings are broadly consistent with prior research findings in this area, expanding upon them and revealing previously unreported aspects of students' thinking. Moreover, our results reflect several new problem contexts in addition to those reported in prior research, and our sample population includes large numbers of both introductory and advanced students. Our detailed findings related to common student difficulties with entropy and the second law of thermodynamics before and after traditional instruction in college physics courses can potentially help instructors of these courses improve student understanding of these concepts. These findings can also be valuable for developing effective research-based curricula and pedagogies to reduce student difficulties and help students develop a solid grasp of these fundamental thermodynamic concepts.

# I. INTRODUCTION AND GOALS OF THE INVESTIGATION

A. Overview

Since many physics courses for science and engineering majors focus on helping students with both conceptual understanding and problem solving, research-based conceptual multiple-choice surveys can be invaluable for investigating students' understanding of physics before and after instruction using various curricula and pedagogies [1-7]. With respect to the specific topic of this paper, we note that prior research suggests that both introductory and upper-level students have many persistent difficulties with introductory thermodynamics concepts [2-40], implying that conceptual surveys in thermodynamics could potentially play a particularly important role in guiding improvements in instruction.

In this paper we carry out an extensive analysis of data obtained through administration of a research-based 78-item multiple-choice survey instrument called the Survey of Thermodynamic Processes and First and Second Laws-Long (STPFaSL-Long). The STPFaSL-Long is a validated survey which was administered to students in traditionally taught undergraduate physics courses at both the introductory and upper-level, as well as to physics graduate students in their first-year, first-semester courses. The details pertaining to the development, validation and administration of the STPFaSL-Long survey can be found elsewhere [41]. Previously, we had investigated student difficulties using the STPFaSL-Short survey containing 33 items [37-39]. In that survey, concepts related to a particular thermodynamic property, e.g., entropy, are combined into a single question with concepts related to other thermodynamics properties, e.g., change in internal energy, work, and heat transfer. This combining can make it challenging to disentangle the difficulties that might be associated with each thermodynamic variable separately. Also, STPFaSL-Short survey items often had answer options that incorporated common alternative conceptions to act as distractors. We note here that, in contrast to STPFaSL-Short, each individual item on STPFaSL-Long is tightly focused on a single concept involving a single thermodynamic variable (such as the net change of entropy in a cyclic process) and the answer options are restricted primarily to "increases, decreases, remains constant," or "larger, smaller, the same," as well as "not enough information." Therefore, the survey data we report here—in which questions related to entropy are clearly separated from questions involving other thermodynamic variables—can be particularly valuable in designing instructional tools that are targeted at students' specific difficulties with entropy and the second law.

Recently, we reported on introductory and upper-level students' difficulties with internal energy, work, and heat transfer based on analysis of data from the STPFaSL-Long survey [40]. Here we focus on students' difficulties with entropy and the second law of thermodynamics at the level characteristic of introductory physics courses. We use data from survey administrations in 12 different traditionally taught classes from four different large universities in the US. In addition to administering the written survey in various courses, we interviewed 11 introductory and 6 upper-level students individually using a think-aloud protocol to get a deeper insight into students' thought processes as they answered the survey questions. Analysis of data for the upper-level students was invaluable in comparing the level of their understanding to that of introductory students. Insights into upper-level students' persistent difficulties on conceptual problems related to entropy and the second law across various contexts can be helpful to instructors both in the advanced thermodynamics and introductory physics courses, offering additional perspective on particularly challenging aspects of the curriculum.

Many conceptual difficulties with entropy and the second law have been documented previously in one form or another [2-40]. The work we report here reproduces many of these previous findings, expands on most of them, and introduces new findings never previously reported; details are in our Results section below. We note that while the data tables in this paper provide quantitative data reflecting students' survey responses, we emphasize the interviews in the narrative since they helped us understand these difficulties much more thoroughly and in greater depth than the survey data alone, offering further insights into some of the conceptual difficulties that had been noted in previous research reports.

B. Prior research

Although there are a number of prior studies [2-40] that have focused on student understanding of thermodynamics, we will mention here only a few recent investigations related specifically to entropy and the second law. Cochran and Heron [19] investigated student understanding of heat engines, finding that students had difficulties with most aspects of the heat engine including relevant variables and processes. In particular, they found that students were often unaware of the connection between limitations on engine efficiency and constraints imposed by the second law of thermodynamics. On a completely different theme, in an investigation targeted on upper-level students, Bucy et al. [23] studied student understanding of entropy in the context of various ideal gas processes. For example, students were asked to consider entropy changes of an ideal gas in both isothermal expansions and free expansions into a vacuum, explaining whether the changes would be positive, negative or zero in each case. They found that students often confuse entropy of the universe with the entropy of the system. Moreover, students had difficulty with the concept of entropy as a state variable. Some of these findings were reproduced in a different context by Christensen et al. [18], who studied students' ideas regarding entropy and the second law in an introductory physics course. This study found that students struggled in distinguishing between entropy of the system, the surroundings, and the "universe" [system + surroundings] and had great difficulty when assessing entropy changes in spontaneous processes, frequently failing to recognize that total entropy would have to increase. In particular, it was found that there was a strong tendency to treat total entropy of the universe as a conserved quantity that would not change during any process. At the same time, even after instruction, many students also asserted that entropy of system or surroundings must increase even in contexts where insufficient information was available to reach such a conclusion. Meltzer [12] investigated upper-level students' understanding of the second law using assessment items similar to those employed by Christensen et al., finding that although a larger proportion of students (compared to the introductory classes) provided correct responses overall, they had perhaps an even stronger tendency to claim that the system entropy must always increase even when that was not the case or it was not possible to infer it from the given context. An investigation by Smith et al. [16] focused on student understanding of entropy, heat engines and the Carnot Cycle, revealing several student difficulties that were similar to those reported in earlier studies. This investigation, as well as that by Christensen et al. [18], also showed that students' understanding was improved by the use of research-based instructional tools. More recently, Loverude [42] reported a nuanced and detailed examination of upper-level students' thinking related to entropy, reflecting the conflict they often experienced between "conservation" reasoning on the one hand and an intuition that entropy must always increase (even in cases where it does not) on the other.

C. Focus of the present study

While we were developing the survey, we wanted to gain some insight into what introductory physics students knew about the relevant thermodynamic concepts *before* they studied them in that course [37]. Therefore, we administered a brief open-ended survey as bonus questions on a midterm exam (for which students obtained extra credit) to students in the first semester of an algebra-based physics course [37]. In that course, the instructor had started discussing thermodynamics, introducing concepts such as temperature, heat capacity, thermal expansion, and heat transfer, but there had not yet been any instruction on the first and second laws of thermodynamics. Students were asked to respond to the following questions:
(1) Describe the first law of thermodynamics in your own words.
(2) Describe the second law of thermodynamics in your own words.
(3) Describe other central or foundational principles of thermodynamics (other than the first and second laws).
Of the 207 students [37], 134 (65%) chose to respond to at least some of these bonus questions. These data, some of which we have previously reported in [37], is repeated here to provide additional perspective on the origins and motivation of the present study.
Of the students who responded to the first law question, 52% stated that energy is conserved, e.g., "Energy cannot be created or destroyed," but only 5% made a statement that included heat transfer as part of this energy conservation law [37]. These responses confirmed that many students in introductory physics have been exposed to the first and second laws of thermodynamics before instruction in the college physics course, implying that it could be useful to administer the survey as a pretest before instruction in introductory courses [37]. Of the students who responded to the second law question [37], 51% gave responses that were at least vaguely related to the second law. (A subset of these responses, i.e., 28%, were reminiscent of an attempt to connect entropy with second law.) Even in these responses from students before instruction, the most common statements about the second law were the well-known overgeneralizations of the law as conveyed by the following statements: "Entropy increases always (and/or everywhere)" or "Energy is always moving toward disarray" (the latter statement suggests confusion between energy and entropy).

These hints of initial confusion about thermodynamic principles even at the beginning of the introductory course lay the grounds for more probing investigation of the evolution of students' thinking over the course of instruction; it is this latter investigation that is reported here. Examples of the prevalence of the difficulties with the second law can be found in both the written responses and the interview data of our primary data sample, described in detail below. In the following, we first describe the methodology for our research followed by presentation of results and discussion on student difficulties with entropy and the second law of thermodynamics. We conclude with a summary and instructional implications.

## II. METHODOLOGY

The Survey of Thermodynamic Processes and First and Second Laws-Long (STPFaSL-Long), a validated survey instrument with 78 items, was used in this research. This instrument focuses on introductory thermodynamics concepts. The details of the development and validation of the STPFaSL-Long survey instrument can be found in Ref. [41] and the survey can be found in [43]. Most items on the survey have four possible answer options; those related to entropy, for example, include "increases," "decreases," "remains the same," and "not enough information." Some of the items (22 out of 78) are true/false (T/F) questions. Throughout this paper, we will discuss 3 sets of data: the written data originating from the validation of the STPFaSL-Long survey instrument, interview data obtained from upper-level students and from introductory algebra-based physics students, and written data with explanations from introductory calculus-based students (on questions administered as a pre-test).

The main survey data we discuss in this paper is from the validation of the STPFaSL-Long. Although we discuss here background information about this data, this is not an exhaustive explanation. The full results can be found in Ref. [41]. In some courses, the survey was administered both before and after instruction in relevant concepts (pre-test and post-test, respectively). In particular, the written data analyzed here were taken by administering the survey in proctored in-person classes as a pre-test (before instruction) and post-test (after students had learned the relevant concepts), but before students' final exam in the course. Students were given some extra credit for completing the survey. These written student data are from 12 different in-person classes from four different large public institutions; students completed the survey in class on Scantrons during a 50-minute class period if they took the entire survey since some students were administered only the first 48 or last 52 questions [41]. We note that there are only 19 distinct problem scenarios in the entire survey; thus, many of the items relate to the same scenario but involve different thermodynamic variables such as internal energy, work, heat transfer, entropy etc. As noted above, the STPFaSL-Long survey was preceded by development of a shorter instrument (STPFaSL-Short) that used a different questioning strategy involving test items of greater complexity and more complex answer options. Students take approximately the same amount of time to do the short and long versions of the STPFaSL survey; students can complete either of the two surveys in a 50-minute class period.

We discuss analysis of student difficulties in the written data from three groups of students (five groups including pre/post-test data from both introductory groups). Students in the "Int-calc" courses were typically engineering majors with some physics, chemistry, and math majors, while students in the "Int-alg" courses were mainly biological science majors and/or those interested in health-related professions. Students included in the upper-level group were typically physics majors in thermodynamics courses, or Ph.D. students in the first semester of their graduate programs who had not taken any graduate-level thermodynamics. Since the survey was administered as a pre-test to the graduate students in the first semester of their Ph.D. program, they were presumed to have taken upper-level undergraduate thermodynamics. A subset of introductory-level students took the survey as both a pre-and post-test (a "matched" group); we found that the performances were identical between the matched and un-matched groups [41]. Unmatched data (i.e., data from all students) are used in this report since the sample size was thus much larger. In particular, since most instructors were reluctant to use two class periods to administer both the pre- and post-test, the matched data included only 214 students in calculus-based introductory physics and 315 students in algebra-based introductory physics. The full sample size including unmatched data is shown in Table I. We also note that not all introductory students answered all survey questions and some introductory recitations were given only the first 48 or last 52 questions to ensure split-test reliability [41]. In Table I, on the post-test (pre-test), out of 492 (753) Int-calc students, 168 (505) were given the first 48 questions, 73 (248) were given the last 52 questions, and 251 (0) were given the full survey. In Table I, on the post-test (pre-test), out of 550 (371) Int-alg students, 170 (173) were given the first 48 questions, 218 (198) were given the last 52 questions, and 162 (0) were given the full survey.

The interview data are from 11 introductory algebra-based physics students and 6 upper-level students from one institution who volunteered following announcement of an opportunity to participate in the study. Each interview lasted between 1-2 hours in one sitting depending upon each student's pace. The interviewed students were given $25 for their participation. The interviews used a semi-structured think-aloud protocol [44, 45]. Students were asked to think-aloud as they answered the survey questions and were not disturbed except being asked to keep talking if they became quiet. Only at the end did we ask them for clarifications of points they had not made clear, particularly if they did not answer correctly. The goal of the interviews was to provide qualitative understanding of the students' reasoning that they employed in selecting answer options on the survey. We did not have an objective of "quantifying" the qualitative data through formal coding and statistical analysis; indeed, the small sample size made this an unrealistic option. The introductory algebra-based physics students who were interviewed were also a part of the validation sample; however, the graduate students who were interviewed were not part of the validation sample.

Finally, 342 students from three Int-calc courses (separate from the 753 students who answered the survey questions on paper Scantrons in class) were asked to answer the survey questions at the beginning of the semester (pre-test) electronically on Qualtrics and provide their reasoning for each question. While many students did not provide meaningful reasoning, some students provided short but informative responses. We will only discuss these written explanations for various survey items for cases in which most students in the interview sample provided only correct responses, such that the interviews in those cases did not provide sufficient insight into reasons for student difficulties. These 342 students were not included in the validation data sample.

## III. RESULTS AND DISCUSSION

The raw student data whose analysis is discussed here were included in raw form in the Appendix of the validation of the STPFaSL-Long paper [41]. The average scores from each group, number of students who participated, and their standard deviations are shown in Table I. The standard deviations provide a basis for assessing the magnitudes of differences in performances of different groups that might be considered meaningful. We have included the distribution of upper-level student scores from Table I in Appendix A as well as the distributions of the post-instruction scores for the introductory groups who were given the full 78-item survey. The post-instruction scores are mildly skewed for both introductory groups, with longer tails above the mean indicating a small proportion of students with very high scores. Scores for upper-level students show a peak at high scores with many students scoring well below that peak.

We note that the "percentage of maximum possible" or POMP-score measure assesses random guessing as follows: for each item, the average POMP score in percentage is given by 100*(average % correct─random guessing %)/(100%─random guessing %) [46, 47]. For questions with four choices, the random guessing percentage would be 25% so if the average score for an item is 25%, the average POMP score would be 0% and if the average score for an item is 75%, the average POMP score would 66.7% etc. For True/False questions, therandom guessing percentage would be 50% so if the average score for an item is 50%, the average POMP score would be 0% and if the average score for an item is 75%, the average POMP score would 50% etc. We note that the tables in this paper report the average scores in each case, not the POMP scores.

TABLE I. Average overall score and standard deviation (SD) of different student groups on the survey along with the number (N) of students who participated in the survey in each group. The post-test performance is shown for the upper-level group and both the pre-test and post-test scores are shown for introductory-level students in the calculus-based (Int-calc) and algebra-based (Int-alg) introductory physics courses. Some students were given the full, 78-item survey and this is shown by (78). Some students were given only the first 48 items or the last 52 items of the 78-item survey, and these groups are represented with (48) and (52) respectively.

| Level | Pre-test (48) | Pre-test (52) | Posttest (78) | Posttest (48) | Posttest (52) |
|---|---|---|---|---|---|
| Upper | … … … | … … … | 76% N = 89 SD = 14% | … … … | … … … |
| Int-calc | 52% N = 505 SD = 9% | 52% N = 248 SD = 10% | 58% N = 251 SD = 11% | 58% N = 168 SD = 11% | 55% N = 73 SD = 12% |
| Int-alg | 51% N = 173 SD = 9% | 51% N = 198 SD = 10% | 52% N = 162 SD = 11% | 55% N = 170 SD = 11% | 58% N = 218 SD = 13% |

To better understand why the pre-post differences are so small, we administered a survey at one university at the beginning of algebra-based and calculus-based introductory physics courses, asking students about their prior knowledge of the First and Second Laws of Thermodynamics. These data show that most students had already been exposed to these concepts either in physics or non-physics courses. (The percentages are higher for the algebra-based courses since students in those courses were primarily juniors and seniors, in contrast to the calculus-based courses in which students were primarily first-year college students.) These data are included in Appendix B.

All the tables in the following sections provide, for all items related to a specific topic, the breakdown for the percentages of correct responses (in boldface), as well as percentage of incorrect responses reflective of specific difficulties. Each section addresses one specific learning difficulty. Note that, due to sample splitting for validation purposes, the number of students who attempted each question in the introductory samples varied widely, as mentioned in Section II, with a post-test range of 324-492 for Int-calc and 332-550 for Int-alg.

A. Difficulties with statements related to the second law of thermodynamics

The survey asks students about the validity of four statements related to the second law of thermodynamics in a true/false format. Table II shows that the level of difficulty of different statements related to the second law varied greatly for introductory students (both in the pre-test and post-test) but was relatively consistent for the upper-level students. Below, we provide some excerpts from student interviews that illustrate these difficulties.

Item 27 was the most challenging of the four questions; it asks students whether the following statement is true or false: "The entropy of any system that cannot exchange energy with its surrounding environment must remain unchanged." Table II shows that while 73% of the upper-level students correctly identified this statement as false, both Int-calc and Int-alg groups' correct-response rates varied from 29%-35% regardless of whether it was on the pre-test or post-test. The low correct-response rates for introductory students even after instruction reflect a widely held student idea that entropy remains unchanged for an isolated system. For example, during an interview, in regard to item 27, one introductory student stated, "That's true 'cause if you're not doing anything to it, you can't really cause a change in entropy." This student further reasoned that if $Q = 0$, then the entropy must remain constant. On the pre-test, some other students explained their reasoning on item 27 as follows: "Entropy can only be influenced by outside forces"; "Entropy is the measure of energy"; "You need a change in temp for the entropy to change"; "It [the system] can't exchange heat so there's no way for it [entropy] to change." Other students provided similar explanations, all of which ignore the fact that the entropy of isolated nonequilibrium systems increases as they more closely approach equilibrium. Our results are consistent with those of Christensen et al. [18] for calculus-based students and Bucy et al. [23] for upper-level students and extend them for the first time to students in algebra-based courses as well.

Item 28 was another challenging true/false question, stated as follows: "No heat engine can be more efficient than a reversible heat engine between given high temperature and low temperature reservoirs." Table II shows that while 83% of the upper-level students correctly identified this statement as true, Int-calc and Int-alg groups' correct-response rates on the post-test were very similar at 65% and 67%, respectively. On the pre-test, it was 56% for the Int-calc group but only 46% for the Int-alg group, suggesting that some learning had occurred in the algebra-based course. Some of the written responses on the pre-test reflected an inclination to link "efficiency" to technology and engineering considerations, rather than to basic physics principles; for example: "Heat engine can be effectively efficient, but the tech does not exist yet"; "There has to be something more efficient out there. How can we be so sure something is the most efficient? The best?"; "It can always be more efficient"; "I do not think you can say all reversible heat engines will be more efficient"; "I am sure there are ways to increase the efficiency of heat transfers without loss of heat due to friction in closed environments." During the interviews, an upper-level student's response suggested that even after instruction, substantial confusion can persist: "I think the Carnot efficiency is the maximum efficiency that you can have….that is what they mean by a reversible heat engine…Actually, reversible heat engine between….No, that's false. If it's reversible, it can't be an engine." These results are consistent with the findings of Cochran and Heron [19] discussed earlier but add considerable perspective due to the technology-related ideas expressed by students during the interviews.

Item 29 is a third true/false statement: "The net heat transfer to the system from a hot reservoir cannot be completely converted to mechanical work in a cyclic process." Although 89% of the upper-level students correctly identified this statement as true, the Int-calc group's high correct-response rate (74%) was statistically identical to their pre-test score, suggesting that their "intuitive" responses fortuitously corresponded to the correct response in this case. By contrast, the Int-alg group did show some gain from pre- to post-test (67% to 77%). Some of the written pre-test responses seemed consistent with the idea that technological means could be used to increase efficiency to an arbitrary degree; for example: "This is possible to achieve if you use the correct process"; "Heat transfers always have the ability to be converted to mechanical work"; "A cyclic process is able to convert the net heat transfer to the system of a hot reservoir"; "You just need a system that can convert it."

TABLE II. Response rates of upper-level, introductory calculus-based (Int-calc), and algebra-based (Int-alg) physics students on true/false questions related to the second law of thermodynamics. All item numbers are marked with an asterisk (*) since they are all T/F questions. Correct responses are in boldface.

| | Item # | Answer Choices | Response rate (%) | | | | |
|---|---|---|---|---|---|---|---|
| | | | Upper Post | Int-calc Post | Int-calc Pre | Int-alg Post | Int-alg Pre |
| **Correct statements related to the Second Law of Thermodynamics** | **27*** | **F** | **73** | **29** | **33** | **35** | **30** |
| | **28*** | **T** | **83** | **65** | **56** | **67** | **46** |
| | **29*** | **T** | **89** | **74** | **73** | **77** | **67** |
| | **35*** | **F** | **84** | **70** | **81** | **74** | **78** |
| Incorrect statements related to the Second Law of Thermodynamics | 27* | T | 27 | 71 | 67 | 65 | 70 |
| | 28* | F | 17 | 35 | 44 | 33 | 54 |
| | 29* | F | 11 | 26 | 27 | 23 | 33 |
| | 35* | T | 16 | 30 | 19 | 26 | 22 |

The fourth true/false statement was item 35: "Since a Carnot engine is a reversible engine, it is 100% efficient." About 80% of the introductory students correctly identified this statement as false on the pre-test, with no (or negative) improvement on the post-test, indicating that most students initially found this statement to conflict with their intuition or expectations. It is notable that many of the *incorrect* explanations given on the pre-test seem to reflect confusion about the meaning of "reversible," for example: "Reversible engines are 100% efficient"; "If it is completely reversable [*sic*], no energy is lost, and the engine is 100% efficient"; "Being reversible seems to mean it can be 100% efficient." In a very different context focusing on quantitative analysis of heat engines, Cochran and Heron [19] also encountered considerable student confusion regarding Carnot engines and second-law constraints on engine efficiency. Smith et al. [16] explored similar questions among students enrolled in upper-level thermodynamics courses but did not extend their investigation to students in introductory courses.

B. Difficulties with entropy as a state variable in the context of cyclic processes

Items 8 and 24 ask students to compare the final entropy with the initial entropy in reversible cyclic processes represented on pressure vs. volume (PV) diagrams (Fig. 1): Is final entropy greater than, less than, or the same as the initial entropy? (The process in the diagram for item 24 is explicitly identified as reversible.) Although most students gave the correct answer "same," nearly a quarter of the introductory students said instead that final entropy would be greater than the initial value. Notably, the proportion of upper-level students giving that same "greater than" response was significantly larger than among the introductory students, as over one-third of upper-level students gave that answer on both items. In fact, Int-alg students had the highest correct-response rate and upper-level students the lowest on both items. As shown in Table III, correct-response rates ranged from 49% to 77%, with the highest rates belonging to the Int-alg students on their pre-test. The response patterns imply that, consistent with some prior research on upper-level students [16, 23], many introductory and advanced students had difficulty with the concept that entropy is a state variable, which must have the same value in identical initial and final states. Moreover, the pattern strongly suggests that difficulties on these items increased after students had had more instruction on the topic, rather than less. A plausible interpretation that would be consistent with results discussed below (in Section C and later) is that students become increasingly comfortable with "entropy increases" arguments as they progress in their physics studies, sometimes to the detriment of their ability to analyze entropy changes in reversible and/or cyclic processes. We note that in the limited context of upper-level thermodynamics courses, Smith et al. [16] found student difficulties with cyclic processes that are similar to those we observed among the introductory students.

The varied reasoning pathways followed by students are illustrated by statements made during the interviews. For example, an introductory student said, in reference to item 8, "…entropy increases because there are more particles inside the system." An upper-level student, referring to item 24, said "Oh wait, it's cyclic, I could have just said reversible just means that entropy increases. So, the final entropy should just be greater…because it's reversible, cyclic." Another upper-level student said, "…the energy due to heat transfer into the system would have to increase…we would have to get heat into the system. And so, the final entropy there would be greater than the initial entropy." A smaller number of students argued during the interviews that entropy would decrease during the cycle, ascribing this decrease either to "compression" or to net positive work being done. (Although both diagrams show no net compression—since initial and final volumes are identical—only item 24 corresponds to net positive work.)

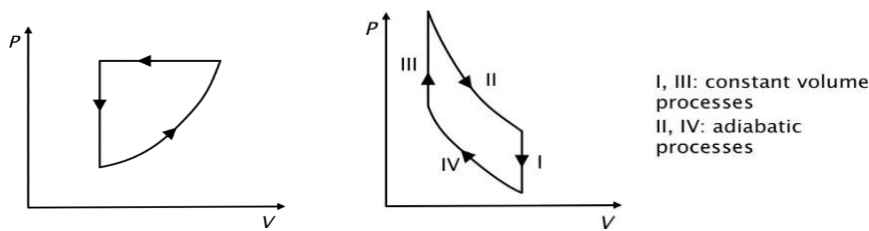

**FIGURE 1.** The diagrams for items 6-9 (left) and items 24-26 (right) on the survey.

TABLE III. Response rates of upper-level, introductory calculus-based (Int-calc), and algebra-based (Int-alg) physics students on questions related to change in entropy of a system after a complete cycle. Answer choice D is not shown because the choice, "not enough information," was not often selected.

| Correct answer in bold | Item # | Answer Choices | Response rate (%) | | | | |
|---|---|---|---|---|---|---|---|
| | | | Upper Post | Int-calc Post | Int-calc Pre | Int-alg Post | Int-alg Pre |
| **$\Delta S_{system} = 0$ after a full cycle (correct)** | **8** | **C** | **49** | **54** | **67** | **65** | **75** |
| | **24** | **A** | **57** | **63** | **63** | **67** | **77** |
| Entropy of a system after a full cycle increases | 8 | A | 36 | 25 | 16 | 19 | 14 |
| | 24 | B | 38 | 27 | 19 | 22 | 16 |
| Entropy of a system after a full cycle decreases | 8 | B | 8 | 16 | 12 | 11 | 8 |
| | 24 | C | 0 | 8 | 13 | 9 | 5 |

C. Difficulties with change in entropy in reversible adiabatic or isothermal processes

The four items 1, 5, 63, and 64 involve entropy changes in adiabatic or isothermal processes which, in all cases, are explicitly identified as being reversible; these items were very challenging for all groups. The relationship $\Delta S = \delta Q_{rev}/T$ (where $Q_{rev}$ is the heat transfer to the system in a reversible process) for reversible processes suffices to answer all four questions (although one may need to apply the first law of thermodynamics to determine the sign of Q for the isothermal processes). Perhaps the most straightforward application of this relationship is in item 1, a reversible adiabatic expansion; since Q = 0, entropy does not change. Nonetheless, all three groups did poorly on this item; post-test correct-response rates were 69%, 47%, and 31%, respectively, for upper level, Int-calc, and Int-alg groups (Table IV). By far the most popular incorrect response was that entropy would increase. An introductory student provided the following argument during an interview: "I think because entropy is like disorder in the system, so I feel like if it expands, disorder may increase, so I guess I will just go with that." Another student followed a different line of thought: "If you're expanding the gas…there is work being done by the gas…I'm just going to go [with entropy] increases." Some of the introductory students' initial ideas are clearly reflected in explanations they offered on the written pre-test questions, for example: "Entropy is always increasing"; "If the gas expands, that means there is more room for the particles to move around, therefore, there can be more disorder or entropy with an adiabatic process"; "The temperature will increase, and since entropy [is] directly proportional to temperature, the entropy will also increase"; "As gas molecules move, more possible arrangements of gas molecules are possible"; "Entropy is the measure of disorder, gas has a lot of disorder and the molecules move very fast and therefore it increases." Some of these seem like reasonable arguments to make before one has explicitly studied the $\Delta S = \delta Q_{rev}/T$ relationship, but even with a significant improvement from pre-test to post-test for the Int-calc students, the post-test score pattern reveals the lingering influence of the pre-instruction ideas. We note that there do not appear to be previous studies of students' ideas regarding entropy changes in adiabatic processes.

The other three survey items involve isothermal processes in which the sign of Q for the system (and thus the sign of the entropy change) depends on whether there is expansion or compression. However, item 64 combines system and surroundings together by asking about "the change in entropy of the gas and the thermal reservoir taken together during the reversible isothermal expansion." Since the (positive) Q for the gas and the (negative) Q for the reservoir must have opposite signs and be equal in magnitude, the $\Delta S = \delta Q_{rev}/T$ relationship again suffices to determine that *total* entropy does not change. The Int-calc group's poor performance on this item was virtually identical to their performance on item 1 and, as was the case for item 1, entropy "increases" was the most popular incorrect answer for all groups both pre- and post-instruction. However, the Int-alg group performed significantly better on this item than they did on item 1, while the upper-level group performed substantially *worse*. The varied initial ideas of the students are again reflected in some of the explanations offered by introductory students on the written pre-test items, for example: "More space = more entropy and thermal reservoir"; "it increases as expansion takes place"; "The expansion causes the increase"; "has to increase because entropy doesn't decrease"; "entropy is always increasing and this is not a closed system."

Item 5 (an isothermal compression for which Q < 0) and item 63 (isothermal expansion, Q > 0) both first require determination of the sign of Q before applying $\Delta S = \delta Q_{rev}/T$ to determine whether entropy of the system decreases or increases. Since it seems—from the student explanations quoted above—that students have an inclination to favor an "entropy increases" idea, one might expect a higher correct-response rate on item 63, for which system entropy actually does increase. Indeed, the performance of the upper-level students is far better on item 63 than it is on the virtually identical item 5 (61% vs. 42% correct). (In a small-sample study of upper-level students [$N = 7$], Bucy et al. [23] observed 100% correct responses regarding the sign of entropy change in an isothermal expansion but did not ask their students to consider isothermal *compressions*; thus, we are unable to make direct comparisons to their work.) However, the performance of the introductory students on item 63 was either identical (on the pre-test) or only marginally better (on the post-test) than their performance on item 5, suggesting that "compression" was, for them, an influential cue that perhaps competed with their inclination to expect entropy increases in all situations. It is also notable that the Int-alg group significantly outperformed the Int-calc group on both items 5 and 63 and even outperformed the upper-level group on item 5. The Int-calc group actually performed significantly *worse* on the post-test (<40% correct) than they did on the pre-test (50%) for both items 5 and 63. Overall, the response that entropy "remains the same" was the most popular incorrect response, while interviews suggested that sign errors were sometimes due to faulty reasoning regarding the sign of heat transfer.

It is interesting to look at students' explanations on item 5, an isothermal compression for which Q < 0 implies that entropy must decrease. The "entropy increases" response was quite popular for this item; an argument advanced during one of the interviews suggests the reasoning involved: "Since it's like a compression, there should be more particles in a smaller space, so entropy should increase." The "remains the same" response was also popular; some explanations include: "It's reversible, then I think entropy might remain the same"; "it would stay the same because entropy is Q over T and T would be constant."

Both "remains the same" and "decreases" were popular incorrect responses on item 63, an isothermal expansion for which system entropy increases. Explanations for the most popular incorrect response "remains the same" included these: "Entropy doesn't change in a reversible one;" "Because it's reversible…the change in entropy of the gas would be zero, so the entropy of the gas would remain the same." It seems that these students were not making a distinction between the system entropy and that of the universe.

TABLE IV. Response rates of upper-level, introductory calculus-based (Int-calc), and algebra-based (Int-alg) physics students on questions related to change in entropy in reversible adiabatic or isothermal processes.

| Correct answer in bold | Item # | Answer Choices | Response rate (%) | | | | |
|---|---|---|---|---|---|---|---|
| | | | Upper Post | Int-calc Post | Int-calc Pre | Int-alg Post | Int-alg Pre |
| *Reversible adiabatic expansion* | | | | | | | |
| **Reversibility (alone) $\Rightarrow \Delta S_{system} = 0$** | **1** | **C** | **69** | **47** | **24** | **31** | **25** |
| $\Delta S_{system} > 0$ | 1 | A | 26 | 37 | 56 | 55 | 57 |
| $\Delta S_{system} < 0$ | 1 | B | 2 | 11 | 17 | 9 | 14 |
| *Reversible isothermal compression* | | | | | | | |
| **$\Delta S_{system} < 0$ (correct)** | **5** | **B** | **42** | **36** | **50** | **48** | **54** |
| $\Delta S_{system} = 0$ | 5 | C | 31 | 33 | 20 | 22 | 22 |
| $\Delta S_{system} > 0$ | 5 | A | 26 | 29 | 28 | 27 | 23 |
| *Reversible Isothermal Expansion* | | | | | | | |
| **$\Delta S_{universe} = 0$ (correct)** | **64** | **A** | **52** | **49** | **48** | **56** | **50** |
| **$\Delta S_{system} > 0$ (correct)** | **63** | **A** | **61** | **39** | **50** | **54** | **55** |
| $\Delta S_{universe} > 0$ | 64 | B | 45 | 28 | 27 | 26 | 27 |
| $\Delta S_{universe} < 0$ | 64 | C | 3 | 19 | 20 | 15 | 17 |
| $\Delta S_{system} = 0$ | 63 | C | 35 | 38 | 25 | 23 | 25 |
| $\Delta S_{system} < 0$ | 63 | B | 4 | 20 | 23 | 22 | 19 |

In general, the response patterns for these four items are consistent with those discussed above in Section B and suggest not only that many students—over the course of instruction—become increasingly inclined to favor an "entropy increases" explanation even in cases where it is not appropriate. It also seems plausible that students in the more "technically advanced" physics courses, such as calculus-based introductory and upper-level, are actually more at risk than are students in the algebra-based courses of developing this often-misleading reasoning pattern during the course of their physics studies. Aside from that, the most consistent theme both in the present study and in those previously reported is persistent student difficulty in distinguishing among entropy of the "system," entropy of the "surroundings," and entropy of the "universe" (system + surroundings). Instructors would be well advised to take special note of this persistent student difficulty.

D. Difficulties with entropy in spontaneous heat transfer between two subsystems

Items 15, 16, and 17 relate to a heat flow process between two solids at different temperatures which are in direct thermal contact with each other, both objects enclosed within an insulated case. The questions relate to the entropy "when thermal equilibrium between the two solids has been established." (Items 16 and 52 ask about the entropy change of the higher-temperature objects, 15 and 51 about the ones at lower temperature, and 17 and 53 about the total entropy of the combined system of both objects.) Although the process as described is not quasi-static and is in any case intrinsically irreversible, one could imagine applying the $\Delta S = \delta Q_{rev}/T$ relationship to the limiting case of infinitesimally slow heat transfer rates. This would then allow one to recognize that entropy of the higher and lower temperature objects would decrease and increase, respectively, while the total entropy of the combined system of two solids would increase, since the entropy increase of the solid initially at the lower temperature would be larger than the entropy decrease of the higher temperature solid. Items 51, 52, and 53 are virtually identical except that instead of two solids in thermal contact, the system consists of "two adjacent identical chambers" containing the same type of gas at different temperatures.

The results (Table V) show that students were very successful in predicting the entropy increases and decreases of the low- and high-temperature objects individually (that is, of the subsystems rather than the combined system); correct-response rates among the introductory students ranged from 69-86% both before and after instruction, with very little improvement from pre-test to post-test ($\leq 8\%$). The upper-level students did somewhat better than the introductory students on "entropy increases" answers for the lower temperature objects, but no better on the "entropy decreases" answers for the higher temperature objects. The introductory students did only marginally better on "entropy increases" answers than they did on "entropy decreases" answers, in contrast to the upper-level students who did much better when the correct answer was "increases." However, the correct-response rates of the introductory students on the "entropy increases" answer for the combined system were *extremely* low, ranging from 20-23%; the most popular incorrect answer by far was that the entropy of the combined system had "not changed." It is notable that the upper-level students performed much better on the combined-system questions, with correct-response rates of 78% and 84% on items 17 and 53, respectively. Overall, the response pattern suggests that introductory students were thinking about the behavior of entropy in a manner similar to how they think about *energy*, such that any energy increase in one subsystem would be balanced by an equal energy *decrease* in the other subsystem. By contrast, most of the upper-level students seem to have come to the realization that entropy and energy behave quite differently.

Many of the statements made during the interviews are very consistent with a confusion between entropy and energy; the students are quite explicit in applying a "conservation of entropy" criterion, for example: "The entropy of the combined system of the two solids has, I feel like no change, kinda like conservation. 'Cause while one is increasing, the other one is decreasing. Like the cold entropy would be increasing"; "I feel like it's overall increase or no change because the entropy lost here [motions to $T_h$] would be gained here [motions to $T_c$]. So I feel like the total entropy wouldn't change." In referring to item 53, a student said, "Since there was no loss of heat from the system or gain of heat by the system, the gain of entropy by the cold reservoir was offset by the loss of entropy of the hot reservoir, so therefore, the entropy is not changed." Other students made very similar arguments: "I think the entropy wouldn't change. It's ... like the internal energy of the combined system, like there is exchange in between, but for entropy, there should be a set amount, it's either positive or negative in the system, but the amount shouldn't change." "I'm going to say that has not changed 'cause if one increases and one decreases, the total is going to not change." By contrast to these clear enunciations of (incorrect) principles, the explanations for incorrect answers on the subsystem questions were more varied and less clear.

We note that some previous studies (e.g., [16] and [23]) have focused on students' understanding of entropy changes related to heat transfer in *reversible* processes, but do not seem to have examined the irreversible contexts that we consider here. Christensen et al. [18] did consider calculus-based students' ideas regarding "spontaneous" processes in which heat transfer occurred, but the problem contexts examined were quite different from those in the present study. Our finding that most students are unaware that net entropy *increases* in irreversible heat-transfer processes is consistent with Christensen et al.'s results but generalizes them substantially, not only to algebra-based and upper-level students, but to physical settings in which the heat transfer is much more explicitly described. Quite similar themes arose in the study reported by Loverude (2015), previously mentioned.

TABLE V. Response rates of upper-level, introductory calculus-based (Int-calc), and algebra-based (Int-alg) physics students on questions related to changes in entropy in spontaneous heat transfer between two subsystems.

| **Correct answer in bold**, | Item # | Answer Choices | Response rate (%) | | | | |
|---|---|---|---|---|---|---|---|
| | | | Upper Post | Int-calc Post | Int-calc Pre | Int-alg Post | Int-Alg Pre |
| *Spontaneous Heat Transfer* | | | | | | | |
| **$\Delta S_{combined\ system} > 0$ (correct)** | **17** | **A** | **78** | **23** | **16** | **20** | **12** |
| | **53** | **A** | **84** | **23** | **14** | **21** | **14** |
| **$\Delta S_{cold} > 0$ (correct)** | **15** | **A** | **97** | **85** | **77** | **86** | **80** |
| | **51** | **A** | **93** | **78** | **72** | **81** | **73** |
| **$\Delta S_{hot} < 0$ (correct)** | **16** | **B** | **74** | **71** | **69** | **80** | **74** |
| | **52** | **B** | **80** | **78** | **70** | **79** | **77** |
| $\Delta S_{combined\ system} = 0$ | 17 | C | 15 | 69 | 75 | 71 | 79 |
| | 53 | C | 11 | 62 | 74 | 68 | 68 |
| $\Delta S_{combined\ system} < 0$ | 17 | B | 5 | 5 | 6 | 7 | 6 |
| | 53 | B | 3 | 10 | 8 | 10 | 13 |
| $\Delta S_{cold} = 0$ | 15 | C | 0 | 6 | 9 | 5 | 10 |
| | 51 | C | 3 | 7 | 12 | 7 | 11 |
| $\Delta S_{cold} < 0$ | 15 | B | 3 | 9 | 11 | 8 | 9 |
| | 51 | B | 3 | 11 | 14 | 12 | 14 |
| $\Delta S_{hot} = 0$ | 16 | C | 6 | 9 | 13 | 8 | 12 |
| | 52 | C | 6 | 6 | 10 | 6 | 6 |
| $\Delta S_{hot} > 0$ | 16 | A | 17 | 17 | 16 | 11 | 13 |
| | 52 | A | 12 | 16 | 17 | 14 | 14 |

E. Difficulties with entropy in free expansion, mixing, or other irreversible processes

The next set of survey items deals with entropy changes in irreversible processes. Item 21 is about a free expansion process (gas expands into larger volume while contained within a thermally insulated container); it is *not* explicitly labeled as irreversible. Similarly, there is no explicit identification of the process as irreversible in items 73, 74, and 75. These items all refer to a constant volume, constant temperature mixing process in which two different gases in separate thermally insulated containers are allowed to mix together. The process referred to in the final set of items (66 and 67) is explicitly identified as "irreversible" in the problem statement; these items refer to entropy changes in an irreversible constant-volume process in which heat is transferred from a thermal reservoir to a gas. Response rates for all of these items are shown in Table VI.

Since all three processes are irreversible, the total entropy of system + surroundings must increase during the process. (In the free expansion process, the gas system is held within an insulated container and thus is isolated from its surroundings.) The items that ask about the entropy changes of the combined system + surroundings are 21 (the gas itself), 67 (the gas + the reservoir), and 75 (the two gases combined, functioning in effect as the other's "surroundings"). The post-instruction correct-response rate of the upper-level students on these items was high: 87%, 83%, and 85% on items 21, 67, and 75, respectively. This was in marked contrast to the introductory students, who did no better than 50% correct on any of the items (range: 35-50% correct on post-test). The Int-alg group did slightly better than the Int-calc group with a 47% average, compared to a 41% average on the three items for the Int-calc group. (The highest correct-response rate for all three groups was on the free-expansion process.) Improvements from pre-test to post-test were modest or nonexistent. The "remains the same" or "not changed" answer was by far the most popular incorrect response on these items overall.

Interview responses make clear that many introductory students were again thinking of entropy as a conserved quantity for which losses or gains in the system would be balanced by gains or losses in the surroundings, for example: "So I think entropy of the system [in the free-expansion process] would remain the same 'cause one of the containers would fill the other one. It would lose the entropy but the other one would gain the entropy. So, I feel like the net would stay the same;" "The heat taken by the gas [in item 67] is coupled with the subsequent loss in heat of the thermal reservoir and if we take those two together, the gain of entropy in the gas should be exactly equal to the loss and thermal energy and heat of the thermal reservoir, so it'll be equal to zero, so it doesn't change." Some of the upper-level students instead justified a "no change" answer on other grounds, for example, the adiabatic nature of the free-expansion process.

TABLE VI. Response rates of upper-level, introductory calculus-based (Int-calc), and algebra-based (Int-alg) physics students on questions related to changes in entropy in free expansion, mixing or other irreversible processes. In this table, we use the subscript "universe" to represent "system + surroundings."

| Correct answer in bold | Item # | Answer Choices | Response rate (%) | | | | |
|---|---|---|---|---|---|---|---|
| | | | Upper Post | Int-calc Post | Int-calc Pre | Int-alg Post | Int-alg Pre |
| *Spontaneous free expansion process of a chemically inert ideal gas* | | | | | | | |
| **$\Delta S_{universe} > 0$ (correct)** | **21** | **B** | **87** | **47** | **43** | **50** | **53** |
| $\Delta S_{universe} = 0$ | 21 | A | 8 | 29 | 30 | 27 | 33 |
| $\Delta S_{universe} < 0$ | 21 | C | 6 | 21 | 24 | 22 | 12 |
| *Spontaneous mixing of two chemically inert ideal gases* | | | | | | | |
| **$\Delta S_{universe} > 0$ (correct)** | **75** | **A** | **85** | **35** | **30** | **43** | **30** |
| **Each $\Delta S_{subsystem} > 0$ (correct)** | **73** | **A** | **79** | **40** | **37** | **48** | **39** |
| | **74** | **A** | **85** | **40** | **38** | **49** | **35** |
| $\Delta S_{universe} = 0$ | 75 | C | 8 | 47 | 54 | 43 | 49 |
| $\Delta S_{universe} < 0$ | 75 | B | 7 | 15 | 14 | 13 | 19 |
| Each $\Delta S_{subsys} = 0$ | 73 | C | 11 | 30 | 30 | 24 | 31 |
| | 74 | C | 11 | 30 | 30 | 24 | 31 |
| Each $\Delta S_{subsys} < 0$ | 73 | B | 9 | 23 | 24 | 23 | 19 |
| | 74 | B | 7 | 24 | 25 | 22 | 25 |
| *Irreversible isochoric process with Q > 0* | | | | | | | |
| **$\Delta S_{universe} > 0$ (correct)** | **67** | **B** | **83** | **40** | **34** | **47** | **41** |
| **$\Delta S_{system} > 0$ (correct)** | **66** | **A** | **88** | **55** | **58** | **58** | **59** |
| $\Delta S_{universe} = 0$ | 67 | A | 14 | 38 | 44 | 34 | 38 |
| $\Delta S_{universe} < 0$ | 67 | C | 2 | 18 | 19 | 17 | 17 |
| $\Delta S_{system} = 0$ | 66 | C | 2 | 20 | 17 | 19 | 18 |
| $\Delta S_{system} < 0$ | 66 | B | 7 | 22 | 22 | 20 | 20 |

The result that the entropy also increases for the gas itself in item 66, and for both types of gases in items 73 and 74, can be arrived at by other arguments, not necessarily as straightforward as in the previous items. For example, in a reversible constant-volume process linking the initial and final states in item 66, heat transfer Q would be greater than zero, thus implying an entropy increase. Similarly, the reversible isothermal expansion linking the initial and final states of the individual gases in items 73 and 74 would involve an increase in system entropy. Correct post-instruction response rates for the introductory students on these items were low for the Int-alg group (48-58%) and even lower for the Int-calc group (40-55%); incorrect responses were fairly closely split between "remains the same" and "decreases." There was little change from pre-test to post-test, although the upper-level group again did significantly better than the introductory students with 79-88% correct responses. Explanations for the "remains the same" responses relied primarily on the fact that temperature did not change (for items 73 and 74) or that volume did not change (for item 66); otherwise, there was little consistency in those responses. Arguments for the "decreases" response varied widely with little consistency.

Students' ideas regarding entropy changes in free-expansion processes were examined by Bucy et al. [18, 23] in a small-sample study of upper-level students (N = 7). After instruction, any remaining confusion seems to have been resolved. We are not aware of any analogous studies in either of the other two problem contexts that we have discussed here.

F.   Difficulties with the law governing the direction of spontaneous processes

Items 20, 56, and 78 are all true/false questions asking whether certain hypothetical "spontaneous" processes would violate the second law of thermodynamics; items 20 and 56 involve heat transfer from low to high temperatures (solids in item 20, gases in item 56), and item 78 involves spontaneous separation of two different gases that have been combined together. The correct answer is "Yes" for all three items. Upper-level students had consistently high correct-response rates on the three items (82-87%; see Table VII). The Int-alg students had nearly identical correct response rates on the two heat-transfer items 20 and 56 (80-79%) but somewhat lower on the gas mixing item 78 (67%). By contrast, Int-calc students did better on the item involving solids (item 20; 77% correct) and worse on the items involving gases (items 56 and 78, 69% and 68% correct). In all cases, pre-test to post-test gains were modest (≤12%) or nonexistent. These results are consistent with prior research which suggests that students do not always connect the second law of thermodynamics with spontaneous processes that occur in nature. For example, one of the explanations offered for item 78 was "I don't think it does. Because you're not changing temperature or anything, so I don't think you're violating either the second or first law."

TABLE VII. Response rates of upper-level, introductory calculus-based (Int-calc), and algebra-based (Int-alg) physics students on questions related to the direction of spontaneous processes. All items are marked with an asterisk (*) since they are T/F questions.

| **Correct answer in bold** | Item # | Answer Choices | Response rate (%) | | | | |
|---|---|---|---|---|---|---|---|
| | | | Upper Post | Int-calc Post | Int-calc Pre | Int-alg Post | Int-alg Pre |
| **Second law of thermodynamics (correct)** | **20*** | **T** | **82** | **77** | **73** | **80** | **68** |
| | **56*** | **T** | **84** | **69** | **70** | **79** | **71** |
| | **78*** | **T** | **87** | **68** | **70** | **67** | **65** |
| First law of thermodynamics | 19* | T | 19 | 53 | 69 | 51 | 68 |
| | 55* | T | 22 | 52 | 63 | 46 | 57 |
| | 77* | T | 21 | 56 | 55 | 52 | 59 |
| Newton's 2nd law | 18* | T | 4 | 17 | 18 | 21 | 28 |
| | 54* | T | 8 | 25 | 24 | 23 | 30 |
| | 76* | T | 10 | 36 | 40 | 34 | 48 |

Other, related items asked whether these same processes would violate the first law of thermodynamics (items 19, 55, and 77) or Newton's 2nd law (items 18, 54, and 76); the correct answer is "No" in all cases. Upper-level students had no difficulty with Newton's second law questions (correct-response rates ≥90%). Most introductory students also answered these items correctly (correct-response rates 64-83%), although performance on the gas-mixing item 76 was somewhat weaker than on the two heat-transfer items.

There were substantially greater difficulties with the questions about the first law of thermodynamics; introductory students' correct-response rates were close to 50% (range: 44%-54%) indicating near-random guessing. Upper-level students did better, with correct responses around 80%. Interviews suggested that many students thought that the heat flow processes would violate the first law principle that "energy is neither created nor destroyed," for example: "Energy cannot be created nor destroyed - this violates that"; "Heat is not created nor destroyed"; "violates the first law of thermodynamics because energy is neither created or destroyed." Other interview responses did not reveal any clear-cut lines of thinking. No one, for example, argued that Newton's second law would be violated if spontaneous processes took place in the opposite direction, although some prior research has suggested that students may think that Newton's second law would be violated because momentum conservation may be violated in the processes involving unmixing of two gases [37-39].

## IV. SUMMARY AND INSTRUCTIONAL IMPLICATIONS

Through analysis of responses on a validated conceptual survey, we studied difficulties related to entropy and the second law of thermodynamics encountered by both introductory and upper-level students after traditional lecture-based instruction. The level of the survey items was that commonly employed in introductory physics courses. The findings presented here suggest that many of these concepts are challenging even for upper-level students, and also suggest that difficulties often persist with little improvement throughout the course of instruction in introductory classes. We find that a widespread unproductive tendency for introductory students to associate the properties of entropy with those of energy leads to many errors based on an idea of "conservation of entropy" in which entropy increases are *always* balanced by equal entropy decreases. For many of the more advanced students (calculus-based and upper level), we detect a tendency to expect entropy increases even in processes in which the entropy does *not* change. We observed a widespread failure to correctly apply the relationship $\Delta S = \delta Q_{rev}/T$, either by using it for *irreversible* processes to which it does not apply, or by applying it incorrectly or completely neglecting it in reversible processes to which it does apply. We also noted that many introductory students are simply not aware that total entropy must increase in any "spontaneous" heat transfer process (that is, from high- to low-temperature). Students at all levels were very frequently found to be confused that while net entropy (system + reservoir) in reversible isothermal processes does not change, the entropy of the working substance itself does indeed increase or decrease depending on whether the process is an expansion or compression. Some of these findings are consistent with and add context to those previously reported in the literature, while others are essentially new results.

Our large and diverse student sample allows us to make comparisons and draw insights that were not possible with previous, more narrowly focused investigations. In particular, our large sample size, wide variety of problem contexts, diversity of student populations, and use of both pre- and post-instruction data as well as interview data greatly increase the reliability and trustworthiness of our key findings. These findings reflect survey data from more than 1000 students from four different universities, with survey questions spanning 19 different problem contexts. The contexts of some of our survey items are the same as those used in previous investigations while other contexts are new. Since the study presented here involves many students from introductory to advanced physics courses using the same survey instrument, it provides further insight into some of the difficulties that have been documented in previous studies which used significantly fewer students and students either from introductory *or* advanced courses but not both. Our findings demonstrate the robustness of some of the previous findings about student difficulties when extended to different student populations employing diverse problem contexts; many examples of this have been discussed in the Results section. A specific example of this is our very consistent finding of a tendency for upper-level students to overgeneralize the "entropy always increases" idea to contexts in which it does not apply. This is a striking validation of similar findings previously reported only for introductory students (e.g., in Christensen et al. [18]) or for very small samples of upper-level students [e.g., Meltzer [12] and Loverude [42]]. The internal consistency of our results and their consistency with prior research reports published over a 20-year period add to our confidence in the reliability of our findings.

In general, our findings are consistent with previously reported results but add substantial breadth and context to them, while opening up some new areas. Our research validates previous findings in new contexts and points to the robustness—or limitations--of previous findings. By extending our study of introductory topics to advanced upper-level students, we gain insights into how and why in certain contexts, the findings in previous studies with students at one level (e.g., introductory or advanced) differ across different levels of students. Through this process, as alluded to above, we have detected substantial evidence for an actual *increase* in student confusion over the course of instruction on some aspects of the entropy concept. Moreover, our finding that student reasoning about these thermodynamics concepts is often strongly context dependent—with specific examples of the context dependence and potential distracting features—provides potentially valuable insights for thermodynamics educators and curriculum developers.

Instructors can utilize the wide variety of problem contexts employed in our survey as a resource to help students develop a functional understanding of relevant concepts that extends beyond a single limited problem type, also helping students develop critical thinking and reasoning skills. This method can be particularly effective when combined with research-based instructional approaches [48]. The findings presented here can be used as baseline data and compared with courses in which innovative evidence-based curricula and pedagogies are used to gauge the level of improvement in introductory and advanced students' understanding of these concepts.

## ACKNOWLEDGMENTS

We thank the students and faculty from all the universities who helped with this research.

## APPENDIX A: DISTRIBUTIONS OF STUDENT OVERALL PERFORMANCE

The distribution of student scores on the entire survey can visually represent the spread of scores seen for each student group. Table I shows the mean performances of each group on the STPFaSL, while Figures 2-4 show the distribution of individual student scores on the entire survey for all in-person groups who were given the full 78-item survey: Upper post, Int-calc post, and Int-alg post.

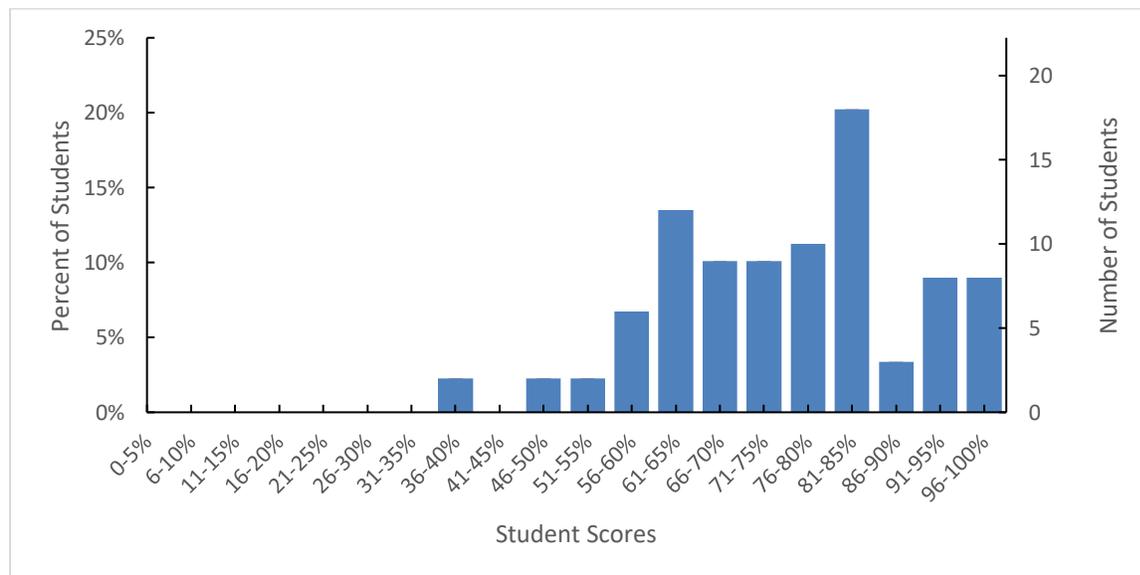

Figure 2: Distribution of student scores for upper post in-person administration. The distribution is shown for the 89 students in this group, binned into 5% increments. The mean score was 76%. The left axis shows the percentage of students that fall into each 5% bin while the right axis shows the number of students that fall into each bin.

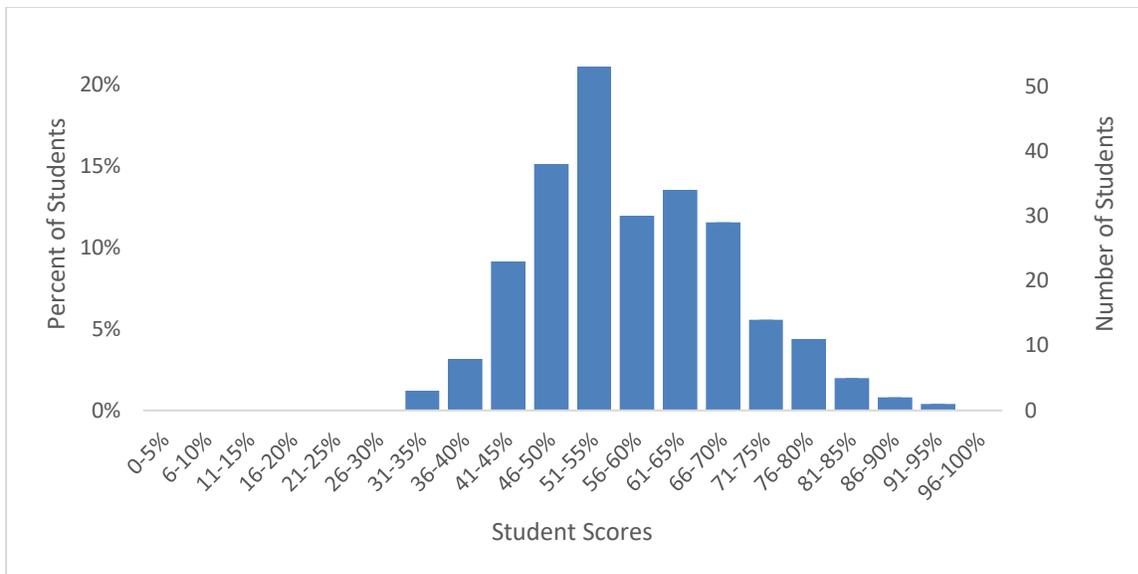

Figure 3: Distribution of student scores for calc-post in-person administration. The distribution is shown for the 251 students in this group, binned into 5% increments. The mean score was 58%. The left axis shows the percentage of students that fall into each 5% bin while the right axis shows the number of students that fall into each bin.

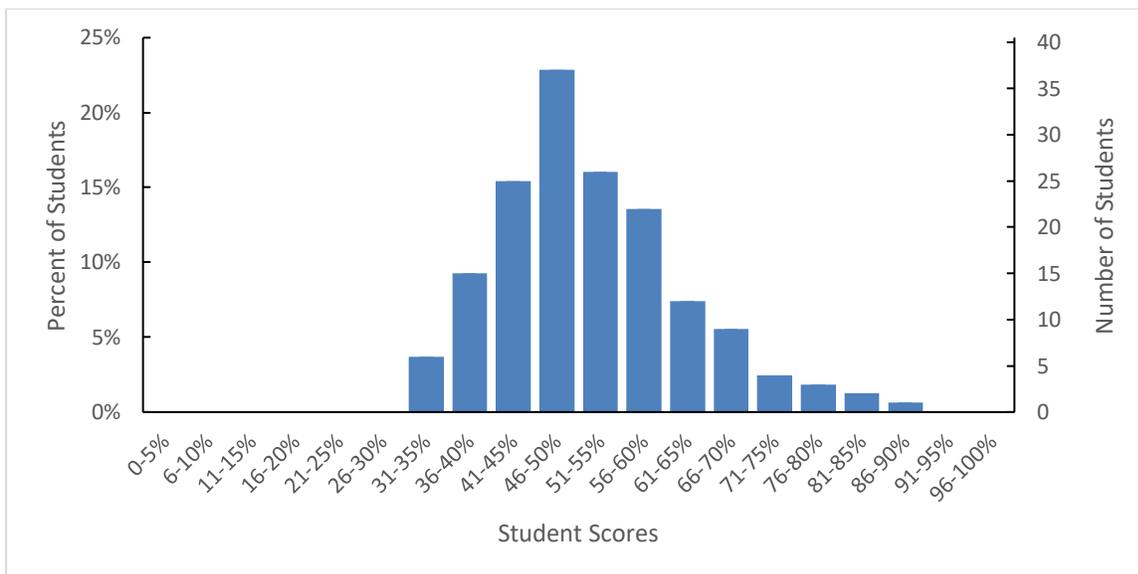

Figure 4: Distribution of student scores for algebra post in-person administration. The distribution is shown for the 162 students in this group, binned into 5% increments. The mean score was 52%. The left axis shows the percentage of students that fall into each 5% bin while the right axis shows the number of students that fall into each bin.

## APPENDIX B: STUDENT PRIOR KNOWLEDGE AT ONE INSTITUTION

To understand why the pre-post differences are generally small, at one university we administered a survey at the beginning of algebra-based and calculus-based introductory physics courses asking students about their prior knowledge of these concepts. The data show that most students have already been exposed to these concepts before (in high school or college) either in a physics or non-physics course. (The percentages are higher for the algebra-based courses since students in these college physics courses were primarily juniors and seniors in contrast to the calculus-based courses in which students were primarily first-year college students).

After the validation of the STPFaSL-Long was done (as reported in ref. [41]), students in a single-semester course were surveyed about their prior knowledge of the First and Second Laws of Thermodynamics. This was done during one semester, at one university, for four different courses: Introductory algebra-based physics 1, algebra-based physics 2, calculus-based physics 1, and calculus-based physics 2. At the beginning of the semester, the students in these four courses were asked if they had seen the First or Second Law previously in a physics course or in another non-physics course. The answer choices were: Yes, No, or Unsure. The two questions were:
- I have studied the first and/or second laws of thermodynamics in a high school or college physics course.
- I have studied the first and/or second laws of thermodynamics in a high school or college chemistry, biology, or other non-physics course.

Incorporating responses to both questions, Tables VIII-XI show whether students have seen the first and/or second laws of thermodynamics in physics classes, non-physics courses, and/or both. The top left percentages in each table below represent students in a course who have studied thermodynamics both in a past physics course and a past non-physics course. For example, for algebra-based physics 1, the first row shows that 60% + 16% + 6% = 82% of the students had seen these concepts in non-physics courses, but 63% of them reported seeing them previously in physics courses. The middle entry in each table represents the percentage of students in each course who had not seen these laws in physics or non-physics courses.

Table VIII: Algebra-based physics 1 students' prior preparation regarding the first and second laws of thermodynamics. The columns show whether students reported studying the first and/or second laws of thermodynamics in a past physics course either in high school or college. The rows show whether students reported studying the first and/or second laws of thermodynamics in a past non-physics course either in high school or college. All percentages in the three-by-three matrix sum to 100%. The number of students self-reporting was N=438.

| Algebra-based Physics 1 | | Physics Course | | |
|---|---|---|---|---|
| | | Yes | No | Unsure |
| Non-Physics Course | Yes | 60% | 16% | 6% |
| | No | 2% | 10% | 0% |
| | Unsure | 1% | 0% | 4% |

Table IX: Algebra-based physics 2 students' prior preparation regarding the first and second laws of thermodynamics. The columns show whether students reported studying the first and/or second laws of thermodynamics in a past physics course either in high school or college. The rows show whether students reported studying the first and/or second laws of thermodynamics in a past non-physics course either in high school or college. All percentages in the three-by-three matrix sum to 100%. The number of students self-reporting was N=86.

| Algebra-based Physics 2 | | Physics Course | | |
|---|---|---|---|---|
| | | Yes | No | Unsure |
| Non-Physics Course | Yes | 80% | 2% | 5% |
| | No | 2% | 3% | 2% |
| | Unsure | 1% | 0% | 3% |

Table X: Calculus-based physics 1 students' prior preparation regarding the first and second laws of thermodynamics. The columns show whether students reported studying the first and/or second laws of thermodynamics in a past physics course either in high school or college. The rows show whether students reported studying the first and/or second laws of thermodynamics in a past non-physics course either in high school or college. All percentages in the three-by-three matrix sum to 100%. The number of students self-reporting was N=656.

| Calculus-based Physics 1 | | Physics Course | | |
|---|---|---|---|---|
| | | Yes | No | Unsure |
| Non-Physics Course | Yes | 38% | 9% | 3% |
| | No | 9% | 25% | 1% |
| | Unsure | 3% | 1% | 12% |

Table XI: Calculus-based physics 2 students' prior preparation regarding the first and second laws of thermodynamics. The columns show whether students reported studying the first and/or second laws of thermodynamics in a past physics course either in high school or college. The rows show whether students reported studying the first and/or second laws of thermodynamics in a past non-physics course either in high school or college. All percentages in the three-by-three matrix sum to 100%. The number of students self-reporting was N=100.

| Calculus-based Physics 2 | | Physics Course | | |
|---|---|---|---|---|
| | | Yes | No | Unsure |
| Non-Physics Course | Yes | 47% | 14% | 4% |
| | No | 4% | 16% | 1% |
| | Unsure | 3% | 0% | 11% |

# References

[1] Heat and Temperature Conceptual Evaluation (HTCE) (2001). https://www.physport.org/assessments/assessment.cfm?A=HTCE.

[2] S. Yeo and M. Zadnik, Introductory thermal concept evaluation: Assessing students' understanding, Phys. Teach. **39**, 496 (2001).

[3] H.-E. Chu, et al., Evaluation of students' understanding of thermal concepts in everyday contexts, Int. J. Sci. Educ. **34**, 1509 (2012).

[4] P. Wattanakasiwich, et al., Development and implementation of a conceptual survey in thermodynamics, Int. J. Innov. Sci. Math. Educ. **21**, 29 (2013).

[5] K. C. Midkiff, T. Litzinger and D. L. Evans, Development of engineering thermodynamics concept inventory instruments, *31st ASEE/IEEE Frontier in Education Conference, Reno, NV.* F2A-F23 (2001) 10.1109/FIE.2001.963691.

[6] R. Streveler, et al., Rigorous methodology for concept inventory development: Using the "assessment triangle" to develop and test the Thermal and Transport Science Concept Inventory (TTCI), Int. J. Eng. Educ. **27**, 968 (2011).

[7] K. D. Rainey, M. Vignal and B. R. Wilcox, Validation of a coupled, multiple response assessment for upper-division thermal physics, Phys. Rev. Phys. Educ. Res. **18**, 020116 (2022).

[8] M. E. Loverude, C. H. Kautz and P. R. L. Heron, Student understanding of the first law of thermodynamics: Relating work to the adiabatic compression of an ideal gas, Am. J. Phys. **70**, 137 (2002).

[9] M. E. Loverude, Student understanding of thermal physics, in *The International Handbook of Physics Education Research: Learning Physics,* edited by M. F. Taşar and P. R. L. Heron (AIP Publishing (online), Melville, New York, 2023), pp. 3.1-3.38.

[10] D. E. Meltzer, Investigation of students' reasoning regarding heat, work, and the first law of thermodynamics in an introductory calculus-based general physics course, Am. J. Phys. **72**, 1432 (2004).

[11] D. E. Meltzer, Investigation of student learning in thermodynamics and implications for instruction in chemistry and engineering, AIP Conf. Proc. **883**, 38 (2007).

[12] D. E. Meltzer, Observations of general learning patterns in an upper-level thermal physics course, AIP Conf. Proc **1179**, 31 (2009).

[13] R. Leinonen, et al., Students' pre-knowledge as a guideline in the teaching of introductory thermal physics at university, Eur. J. Phys. **30**, 593 (2009).

[14] R. Leinonen, M. A. Asikainen and P. E. Hirvonen, University students explaining adiabatic compression of an ideal gas - a new phenomenon in introductory thermal physics, Res. Sci. Educ. **42**, 1165 (2012).

[15] R. Leinonen, M. A. Asikainen and P. E. Hirvonen, Overcoming students' misconceptions concerning thermal physics with the aid of hints and peer interaction during a lecture course, Phys. Rev. ST Phys. Educ. Res. **9**, 020112 (2013).

[16] T. I. Smith, et al., Identifying student difficulties with entropy, heat engines, and the Carnot cycle, Phys. Rev. ST Phys. Educ. Res. **11**, 020116 (2015).

[17] B. W. Dreyfus, et al., Resource Letter TTSM-1: Teaching thermodynamics and statistical mechanics in introductory physics, chemistry, and biology, Am. J. Phys. **83**, 5 (2015).

[18] W. M. Christensen, D. E. Meltzer and C. A. Ogilvie, Student ideas regarding entropy and the second law of thermodynamics in an introductory physics course, Am. J. Phys. **77**, 907 (2009).

[19] M. J. Cochran and P. R. L. Heron, Development and assessment of research-based tutorials on heat engines and the second law of thermodynamics, Am. J. Phys. **74**, 734 (2006).

[20] J. M. Bennett and M. Sözbilir, A study of Turkish chemistry undergraduates' understanding of entropy, J. Chem. Educ. **84**, 1204 (2007).

[21] H. Georgiou and M. D. Sharma, Does using active learning in thermodynamics lectures improve students' conceptual understanding and learning experiences?, Eur. J. Phys. **36**, 015020 (2014).

[22] K. Bain, et al., A review of research on the teaching and learning of thermodynamics at the university level, Chem. Educ. Res. Pract. **15**, 320 (2014).

[23] B. R. Bucy, J. R. Thompson and D. B. Mountcastle, What is entropy? Advanced undergraduate performance comparing ideal gas processes, AIP Conf. Proc. **818**, 81 (2006).

[24] J. W. Clark, J. R. Thompson and D. B. Mountcastle, Comparing student conceptual understanding of thermodynamics in physics and engineering, *AIP Conf. Proc.* **1513***,* 102-105 (2013) https://aip.scitation.org/doi/abs/10.1063/1.4789662.

[25] H. Goldring and J. Osborne, Students' difficulties with energy and related concepts, Phys. Educ. **29**, 26 (1994).

[26] M. F. Granville, Student misconceptions in thermodynamics, J. Chem. Educ. **62**, 847 (1985).

[27] C. H. Kautz, et al., Student understanding of the ideal gas law, Part I: A macroscopic perspective, Am. J. Phys. **73**, 1055 (2005).

[28] C. H. Kautz and G. Schmitz, Probing student understanding of basic concepts and principles in introductory engineering

thermodynamics, *Proceedings of the ASME 2007 International Mechanical Engineering Congress and Exposition. Volume 6: Energy Systems: Analysis, Thermodynamics and Sustainability.* 473-480. (2007) https://doi.org/10.1115/IMECE2007-41863.

[29] E. Langbeheim, et al., Evolution in students' understanding of thermal physics with increasing complexity, Phys. Rev. ST Phys. Educ. Res. **9**, 020117 (2013).

[30] M. Malgieri, et al., Improving the connection between the microscopic and macroscopic approaches to thermodynamics in high school, Phys. Educ. **51**, 065010 (2016).

[31] T. Nilsson and H. Niedderer, An analytical tool to determine undergraduate students' use of volume and pressure when describing expansion work and technical work, Chem. Educ. Res. Pract. **13**, 348 (2012).

[32] T. I. Smith, D. B. Mountcastle and J. R. Thompson, Identifying student difficulties with conflicting ideas in statistical mechanics, *AIP Conf. Proc.* **1513**, 386-389 (2013) https://aip.scitation.org/doi/abs/10.1063/1.4789733.

[33] T. I. Smith, D. Mountcastle and J. Thompson, Student understanding of the Boltzmann factor, Phys. Rev. ST Phys. Educ. Res. **11**, 020123 (2015).

[34] P. L. Thomas and R. W. Schwenz, College physical chemistry students' conceptions of equilibrium and fundamental thermodynamics, Journal of Research in Science Teaching **35**, 1151 (1998).

[35] P. V. Engelhardt, An introduction to Classical Test Theory as applied to conceptual multiple-choice tests, in *Getting Started in PER*, edited by C. Henderson and K. Harper (AAPT, College Park, MD, 2009), Reviews in PER, Vol. 2.

[36] B. Brown, Developing and Assessing Research-Based Tools for Teaching Quantum Mechanics and Thermodynamcis, PhD Dissertation, University of Pittsburgh, 2015.

[37] B. Brown and C. Singh, Development and validation of a conceptual survey instrument to evaluate students' understanding of thermodynamics, Phys. Rev. Phys. Educ. Res. **17**, 010104 (2021).

[38] B. Brown and C. Singh, Student understanding of the first law and second law of thermodynamics, Eur. J. Phys. **42**, 065702 (2021).

[39] B. Brown and C. Singh, Student understanding of thermodynamic processes, variables and systems, Eur. J. Phys. **43**, 055705 (2022).

[40] M. J. Brundage, D. E. Meltzer and C. Singh, Investigating introductory and advanced students' difficulties with change in internal energy, work, and heat transfer using a validated instrument, Phys. Rev. Phys. Educ. Res. **20**, 010115 (2024).

[41] M. J. Brundage and C. Singh, Development and validation of a conceptual multiple-choice survey instrument to assess student understanding of introductory thermodynamics, Phys. Rev. Phys. Educ. Res. **19**, 020112 (2023).

[42] M. Loverude, Identifying student resources in reasoning about entropy and the approach to thermal equilibrium, Phys. Rev. ST Phys. Educ. Res. **11**, 020118 (2015).

[43] PhysPORT STPFaSL-Long survey instrument, https://www.physport.org/assessments/assessment.cfm?A=STPFaSLlong.

[44] V. K. Otero, D. B. Harlow and D. E. Meltzer, Qualitative methods in physics education research, in *The International Handbook of Physics Education Research: Special Topics,* edited by M. F. Taşar and P. R. L. Heron (AIP Publishing (online), Melville, New York, 2023), pp. 25.1-25.32.

[45] K. A. Ericsson and H. A. Simon, *Protocol analysis; Verbal reports as data (revised edition). Bradfordbooks,* (MIT Press, Cambridge, MA, 1993).

[46] P. Cohen, et al., The Problem of Units and the Circumstance for POMP, Multivariate Behavioral Research **34**, 315 (1999).

[47] R. Sayer, E. Marshman and C. Singh, Case study evaluating Just-In-Time Teaching and peer instruction using clickers in a quantum mechanics course, Phys. Rev. Phys. Educ. Res. **12**, 020133 (2016).

[48] Tutorials in Thermal & Statistical Physics, https://www.physport.org/methods/method.cfm?G=Thermal_Tutorials.